\documentclass[aps,pra,twocolumn,superscriptaddress,amsmath,amssymb]{revtex4-1}
\usepackage{natbib}
\usepackage{graphics}
\usepackage{graphicx}
\usepackage{color}
\usepackage{verbatim}
\usepackage{hyperref}
\usepackage{amsmath}
\usepackage{bm}
\usepackage{lipsum}

\renewcommand{\r}{{\bm r}}
\newcommand{\p}{{\bm p}}

  \newcommand{\n}{{\bm n}}

  \newcommand{\eref}[1] {(\ref{#1})}
\newcommand{\Eref}[1] {Eq.~(\ref{#1})}

\begin{document}

% Use the \preprint command to place your local institutional report
% number in the upper righthand corner of the title page in preprint mode.
% Multiple \preprint commands are allowed.
% Use the 'preprintnumbers' class option to override journal defaults
% to display numbers if necessary
%\preprint{}

%Title of paper
\title{Strong field ionisation of Argon: Electron momentum spectra and nondipole effects}

% repeat the \author .. \affiliation  etc. as needed
% \email, \thanks, \homepage, \altaffiliation all apply to the current
% author. Explanatory text should go in the []'s, actual e-mail
% address or url should go in the {}'s for \email and \homepage.
% Please use the appropriate macro foreach each type of information

% \affiliation command applies to all authors since the last
% \affiliation command. The \affiliation command should follow the
% other information
% \affiliation can be followed by \email, \homepage, \thanks as well.
%\author{}
%\email[]{}
%\homepage[]{Your web page}
%\thanks{}
%\altaffiliation{}
\author{Nida Haram}
\email{nida.haram@griffithuni.edu.au}
\affiliation{Centre for Quantum Dynamics, Griffith University, Brisbane, Queensland 4111, Australia}

\author{Han Xu}
%\email{h.xu@griffith.edu.au}
\affiliation{Centre for Quantum Dynamics, Griffith University, Brisbane, Queensland 4111, Australia}

\author{Igor Ivanov}
%\email{igorivanov@ibs.re.kr}
\affiliation{Centre for Relativistic Laser Science, Institute for Basic Science, Gwangju 61005, Republic of Korea}

\author{D. Chetty}
\affiliation{Centre for Quantum Dynamics, Griffith University, Brisbane, Queensland 4111, Australia}

\author{Igor Litvinyuk}
\email{i.litvinyuk@griffith.edu.au}
\affiliation{Centre for Quantum Dynamics, Griffith University, Brisbane, Queensland 4111, Australia}

\author{R.T. Sang}
\email{r.sang@griffith.edu.au}
\affiliation{Centre for Quantum Dynamics, Griffith University, Brisbane, Queensland 4111, Australia}

%Collaboration name if desired (requires use of superscriptaddress
%option in \documentclass). \noaffiliation is required (may also be
%used with the \author command).
%\collaboration can be followed by \email, \homepage, \thanks as well.
%\collaboration{}
%\noaffiliation

\date{\today}

\begin{abstract}% insert abstract here
We investigate the influence of relativistic nondipole effects on the  photoelectron spectra of argon, particularly in the low kinetic energy region (0 eV - 5 eV). In our experiment, we use intense linearly polarised 800 nm laser pulse to ionise Ar from a jet and we record photoelectron energy and momentum distributions using a reaction microscope (REMI). Our measurements show that nondipole effect can cause an energy dependent asymmetry along the laser propagation direction in the photoelectron energy and momentum spectra. Model simulation based on time-dependent Dirac equation (TDDE) can reproduce our measurement results. The electron trajectory analysis based on classical model reveals that the photoelectron which obtains negative momentum shift along laser propagation direction is caused by the interplay between the Lorenz force induced radiation pressure during its free propagation in continuum and re-scattering by Coulomb potential of the parent ion when it is driven back by the laser field. 

%{\color{red}whatever}
\end{abstract}

% insert suggested PACS numbers in braces on next line
%\pacs{}
% insert suggested keywords - APS authors don't need to do this
%\keywords{}

%\maketitle must follow title, authors, abstract, \pacs, and \keywords
\maketitle

% body of paper here - Use proper section commands
% References should be done using the \cite, \ref, and \label commands
%\section{Introduction}% Put \label in argument of \section for cross-referencing%\section{\label{}}

\section{Introduction}

In strong field physics, simple man’s three step model \cite{Corkum_1993} has been proven to be extremely useful for providing quantitative explanations to a wealth of highly nonlinear laser-matter interaction phenomena, such as high-order harmonic generation \cite{McPherson_1987,Ferray_1988,Walser_2000}, above-threshold ionisation (ATI) \cite{Corkum_1989,Eberly_1991,Milosevic_2006}, non-sequenctial double ionisation \cite{Fittinghoff_1992,Walker_1994,Feuerstein_2001}, frustrated tunneling ionisation \cite{Nubbermeyer_2008}, etc. In this semiclassical model, the laser-matter interaction happens in two main stages. At the first stage, the target atom or molecule is tunnel ionised and a free electron appears in the continuum. At the second stage, the free electron is driven solely by laser's electric field while the influence of the Coulomb potential is neglected.

%.......<breakdown of simple man’s model>.......

%<Short and long range Coulomb effect>
The predictions of simple man’s model, however, could not show satisfactory quantitative agreement with the results of high-resolution strong field experiments. For example, the existence of low energy structure (LES), very low energy structure and near zero energy structures in the photoelectron energy and momentum spectra obtained as a result of strong field ionisation with low frequency fields \cite{Blaga_2012, Faisal_2009,Wu_2012,Quan_2009,Dura_2013, Pullen_2014,Wolter_2014,Wolter_2015} that arise due to the short-range Coulomb focusing effect, which is an interaction that is ignored in simple man’s model. Such short-range Coulomb focusing effects can only be observed when the free electron is driven back to its parent ion core with a distance from several tens to 100 atomic units with nearly zero velocity. Since the Coulomb force scales inversely to the distance from the core (U $\sim$ 1/r), the long-range Coulomb effect is supposed to be much weaker.  The long-range Coulomb effect has been found to significantly affect the emission angle of the streaked photoelectrons \cite{willenberg_2019a} and strongly influence the zero energy structure of the photoelectron momentum spectrum \cite{Quan_2016}.

The suppression of LES in case of experiment  with circularly polarised laser field led to the speculation that their origin might be linked with rescattering \cite{Faisal_2009}. In accordance with the results from circularly polarised laser fields, the time-dependent Schr\"{o}dinger equation (TDSE) results also suggested that rescattering is involved in the creation of LES.  Initial theoretical investigations \cite{Yan_2010,Liu_2010} attributed LES to be the result of interplay between forward scattering and Coulomb focusing effect, which supported the proposition of recollision. Another model suggested that the creation of LES is linked with the bunching of photoelectrons, which miss the parent ion upon the so-called soft recollisions \cite{Kastner_2012a,Kastner_2012b}. The insight into these low energy features is of great importance for photoelectron holography to determine the structural properties of the parent ion \cite{Meckel_2008, Bian_2012, Meckel_2014} and to explore the time-resolved tunneling dynamics \cite{Porat_2018, willenberg_2019a}.

%...........<Main aim of our project>...........
In this work, we investigate the influence of relativistic nondipole effects on the low energy features of the photoelectron momentum and energy spectra obtained by using moderately intense linearly polarised near-infrared laser field. These effects become discernible in the form of an offset in the photoelectron momentum distribution along the laser propagation direction \cite{Smeenk_2011,Ludwig_2014,Ivanov_2016,Haram_2019}. The relativistic nondipole effects encompass the short-range Coulomb focusing effect, demonstrating the breakdown of strong field approximation and also of dipole approximation. In the former approximation, the Coulomb interaction of the photoelectron with the parent ion after the ionisation is neglected \cite{Marco_1992,Ivanov_2006}. The later approximation assumes that the electromagnetic field is spatially uniform on the length scale of the electron motion during the laser pulse and the influence of magnetic field component of the laser pulse on the electron motion is negligible compared to that of the electric field component \cite{Reiss_1992,Haram_2020}. Breakdown of these approximations gives rise to the relativistic nondipole effects which induce non-negligible momentum transfer to the photoelectrons, which ultimately modifies the low energy features of the photoelectron momentum and energy spectra.

We present high resolution, ultra-low energy photoelectron momentum and energy spectra for the single ionisation of Ar by few-cycle, linearly polarised laser pulses (800 nm, 6 fs, 0.8 PW/cm$^{2}$) under exceptionally clean experimental conditions. We check normalised asymmetry of the low energy features in the laser propagation direction to study the impact of relativistic nondipole effects on them. We show that the strong kinetic energy dependence of the asymmetry is caused by the ionisation phase and thus by the trajectory of the tunneled electron. For slow electrons, with their kinetic energy close to zero, the propagation trajectory in the continuum allows these electrons to experience strong Coulomb focusing. We show that such strong interaction can cause a very interesting kinetic energy dependent peak shifting and spectral narrowing on the transverse electron momentum distribution (TEMD). The experimental results are supported by the truly relativistic ab initio three dimensional time dependent Dirac equation (3D-TDDE) \cite{Haram_2019}.

\section{Experimental System}
\begin{figure}[h!]
\includegraphics[scale=0.35]{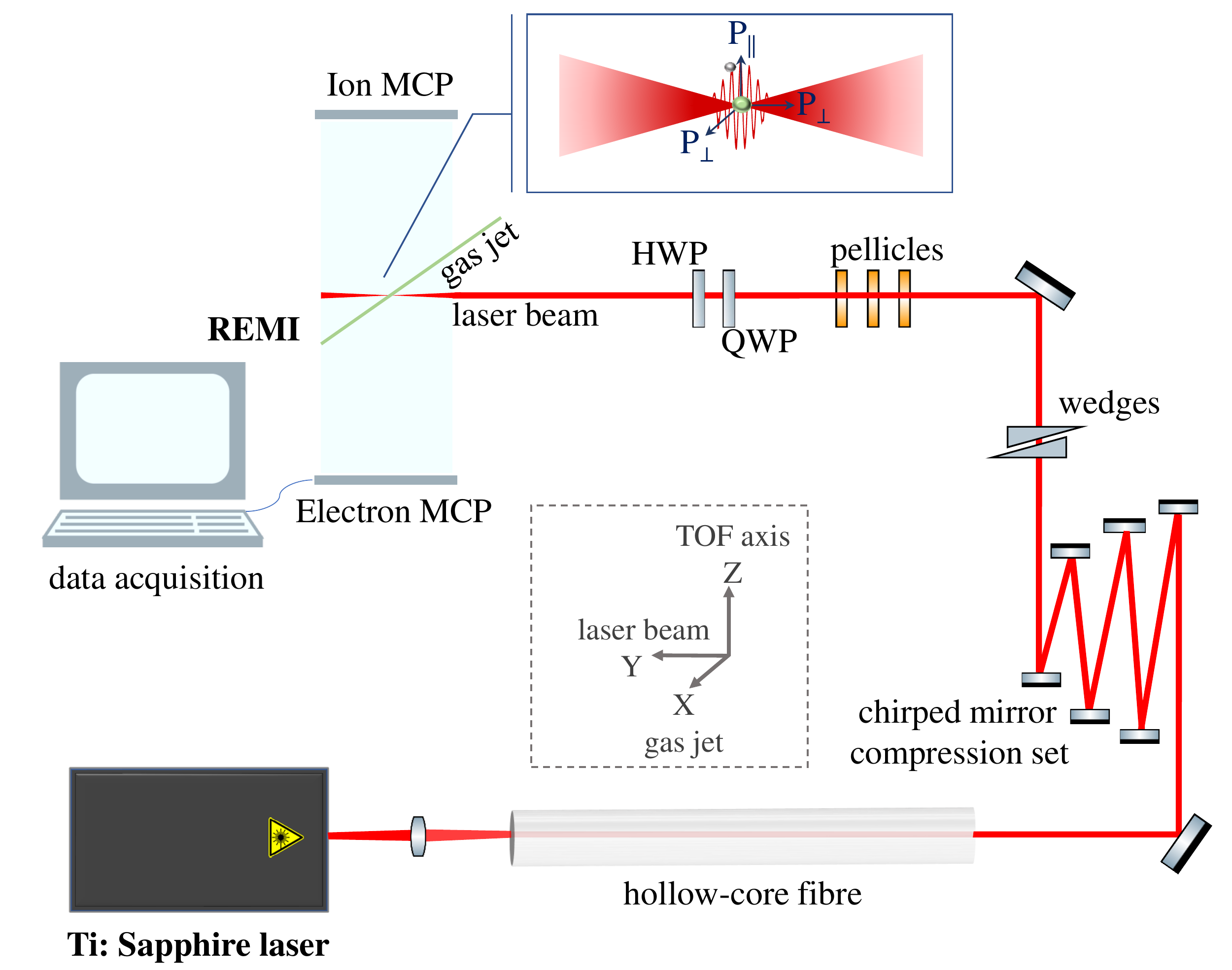}%
\caption{\label{fig:expsetup}}The experimental setup used to explore the influence of relativistic nondipole effects on the low energy features of momentum and kinetic energy spectra 
\end{figure}

%..............<Laser system>...........
The experiments were performed with a commercially available Ti:sapphire laser system in combination with a reaction microscope (REMI). The laser system delivers linearly polarised few cycle pulses in the near infrared (central wavelength 800 nm) at 1 kHz repetition rate. The inherent ellipticity of the laser pulses was eliminated by a quarter wave plate (QWP). The pulse duration in the reaction zone was controlled by compensating the chirp using a pair of fused silica wedges. The laser beam was tightly focused to a spot size of 7.25 $\mu$m in the reaction zone inside the ultrahigh vacuum chamber (10$^{-10}$ mbar) by a silver coated spherical mirror of focal length 75 mm. The laser intensity was controlled by using pellicle beamsplitters and was chosen such that only single ionisation of Ar remains dominant.  The laser beam along y-axis was crossed by the supersonic gas jet along x-axis at the laser focus from where the resulting ions and electrons travel along z-axis/time of flight (TOF) axis to their respective detectors of the REMI (see Fig. \ref{fig:expsetup}).  A half-wave plate (HWP) was used to rotate the polarisation axis of the laser pulses to TOF axis of the REMI. The laser intensity in the reaction zone was determined precisely by the recoil-ion momentum imaging method within 10\% confidence interval \cite{Alnaser_2004,Smeenk_2011_1}. To guarantee the reliability of our data, laser intensity was constantly monitored during the experiment with intensity fluctuations less than 5\%.

%.............<reaction microscope>...........
The complete final state momentum and energy distributions of both electrons and ions were recorded by REMI \cite{Ullrich_2003} with a controlled well-defined momentum resolution. To avoid any obstruction in the electron spectra, coincident detection of Ar$^{+}$ ion allows to avoid any contributions of electrons coming from higher ionisation states as we consider only those electrons which conserve momentum with Ar$^{+}$ ion. Moreover, the space charge effects were avoided by carefully adjusting the laser intensity and width of the supersonic gas jet in the reaction zone so that it does not influence the final electron momentum distribution. A very low electric field (55 V/cm) was applied parallel to the TOF axis and laser polarisation but perpendicular to the laser beam and supersonic gas jet direction, which guides the low energy electrons and ions towards two time and position sensitive micro-channel (MCP) detectors. A homogeneous weak magnetic field (1.95 G) was applied to confine the transverse motion of the electrons to ensure that the electrons with energies less than 30 eV reach the detector. These low values of electric and magnetic field improved the resolution of REMI such that we were able to observe the low energy features with high resolution and clarity. \\

Measurements were taken at two different laser intensities for comparison purpose: 0.3 PW/cm$^{2}$ and 0.8 PW/cm$^{2}$. The value for the low intensity was chosen such that the relativistic effects are least prominent (refer to our previously published paper \cite{Haram_2019}) but can give us sufficient count rate. The maximum laser intensity was selected so as to avoid obstruction in the photoelectron momentum spectra as a result of the space charge effect or multiple ionisation of the gas species.

\section{Theory}
\subsection{Theoretical model based on 3D-TDDE}

To solve the TDDE we follow the procedure described in \cite{tdde_theory1,tdde_theory_2,crotm1,Haram_2019}, which
we briefly recapitulate here. We solve the TDDE for the Ar atom in the field of a laser pulse described by the vector potential $A_z(t-y/c)$ (the pulse is linearly polarised in $z-$ direction and propagates in $y-$ direction): 

\begin{equation}
i{ \frac{\partial \Psi(\r,t)}{\partial t} }={\hat H} \Psi(\r,t) \ ,
\label{tdde1}    
\end{equation}

\noindent where $\Psi(\r,t)$ is a four-component wave-function (bispinor), ${\hat H}$ the Hamiltonian operator:

\begin{equation}
\hat H = \hat H_{\rm atom}+ \hat H_{\rm int}
\label{tdde2}\ ,
\end{equation}

\noindent where:

\begin{equation}
\hat H_{\rm atom}=c{\bm \alpha}\cdot {\hat{\bm p}}+
c^2(\beta-I)+ I V(r) ,
\label{tdde3}
\end{equation}

\noindent and

\begin{equation}
\hat H_{\rm int}=c{\bm \alpha}\cdot {\bm A}(t,y) \ ,
\label{tdde4}
\end{equation}

\noindent where $c=137.036 \; a.u.$- the speed of light in atomic units. We use the Dirac basis for the $\bm \alpha-$ matrices: 
$\displaystyle {\bm \alpha}=\left( \begin{array}{cc}
{\bm 0} & {\bm \sigma} \\
 {\bm \sigma} & {\bm 0}  \\
 \end{array} \right)$, $\displaystyle \beta=\left( \begin{array}{cc}
{\bm I} & {\bm 0} \\
{\bm 0} & -{\bm I}  \\
 \end{array} \right)$, $\displaystyle I=\left( \begin{array}{cc}
{\bm I} & {\bm 0} \\
{\bm 0} & {\bm I}  \\
 \end{array} \right)$, ${\bm \sigma}$ are Pauli matrices,
${\bm 0}$ and ${\bm I}$ are $2\times 2$ null and identity matrices. We subtracted from the field-free atomic Hamiltonian \eref{tdde3} the rest mass term $Ic^2$ so that we operate in a more familiar energy scale commonly used in the non-relativistic atomic physics calculations.
Ar atom is described in the single active electron (SAE) approximation, we use for $V(r)$ in \Eref{tdde3}  the model potential  given in \cite{tdde_theory_3}.

Solution is represented as a series in basis bispinors:

\begin{equation}
\Psi({\bm r},t)=
\sum\limits^{J_{\rm max}}_{j\atop l=j\pm 1/2} \sum\limits_{M=-j}^{j} 
\Psi_{jlM}({\bm r},t),
\label{basis}
\end{equation}

\noindent where each basis bispinor is:

\begin{equation}
\Psi_{jlM}({\bm r},t)=
\left( \begin{array}{c}
g_{jlM}(r,t)\Omega_{jlM}(\n) \\
f_{jlM}(r,t)\Omega_{jl'M}(\n)  \\
 \end{array} \right),
\label{bb}
\end{equation}

\noindent and two-component spherical spinors are defined as $\displaystyle \Omega_{jlM}(\n)=
\left( \begin{array}{c} 
C^{jM}_{l\ M-{\frac{1}{2}} {\frac{1}{2}}{\frac{1}{2}}}Y_{l,M-{\frac{1}{2}}}(\n) \\
C^{jM}_{l\ M+{\frac{1}{2}} {\frac{1}{2}}-{\frac{1}{2}}}Y_{l,M+{\frac{1}{2}}}(\n) 
\end{array} \right)$,

\noindent (here $C^{jM}_{lm{\frac{1}{2}}\mu}$ are 
the Clebsch-Gordan coefficients, $Y_{lm}(\n)$- the spherical harmonics, and $\n=\r/r$). Parameters $l$ and $l'$ in  \Eref{basis} must satisfy the relation $l+l'=2j$.

To take into account the non-dipole effects due to the spatial dependence of the laser fields, vector potential in \Eref{tdde4} is expanded in a series of spherical harmonics at every step of the integration procedure. Substituting expansion \eref{basis} in TDDE \eref{tdde1} and performing some cumbersome but straightforward manipulations using well-known properties of spherical spinors \cite{ll4}, one obtains a set of the coupled differential equations for the radial functions  $g_{jlM}(r,t)$ and $f_{jlM}(r,t)$ in \Eref{bb}. This system is solved using a  generalisation of the well-known matrix iteration method (MIM) \cite{velocity1}, which we described in detail in \cite{tdde_theory1}. 

Differential ionisation probabilities are calculated as $P({\mu},\p)=|a({\mu},\p)|^2$, where $a({\mu},\p)$ are the ionisation amplitudes (here $\mu$ is polarisation (i.e., spin direction in the electron's rest frame), and $\p$ asymptotic electron momentum). The amplitudes are obtained by projecting solution of the TDDE after the end of the pulse on the set of the ingoing relativistic scattering states $\Psi^-_{\mu,\p}(\r)$ of Ar atom calculated numerically using  the model potential $V(r)$ in \Eref{tdde3}. As we are not able to resolve different electron spin states in the experiment, we sum our results for the probability distributions over electron polarisation states.

\section{Results and Discussion}
The photoelectron momentum perpendicular to the laser polarisation plane, also known as transverse momentum, carries the information about the recollision event caused by Coulomb focusing of the photoelectrons  \cite{Corkum1993,Brabec1996,Blaga2009,Quan2009,Korneev2012,Liu2012} without being disturbed by the momentum transfer from the laser field. Therefore, TEMD is considered as the most suitable spectrum to explore the influence of relativistic nondipole effects on the low energy features. In addition, the dependence of transverse momentum on the kinetic energy of photoelectrons provides more insight into the underlying physics.

\begin{figure}[h!]
\includegraphics[scale=0.51]{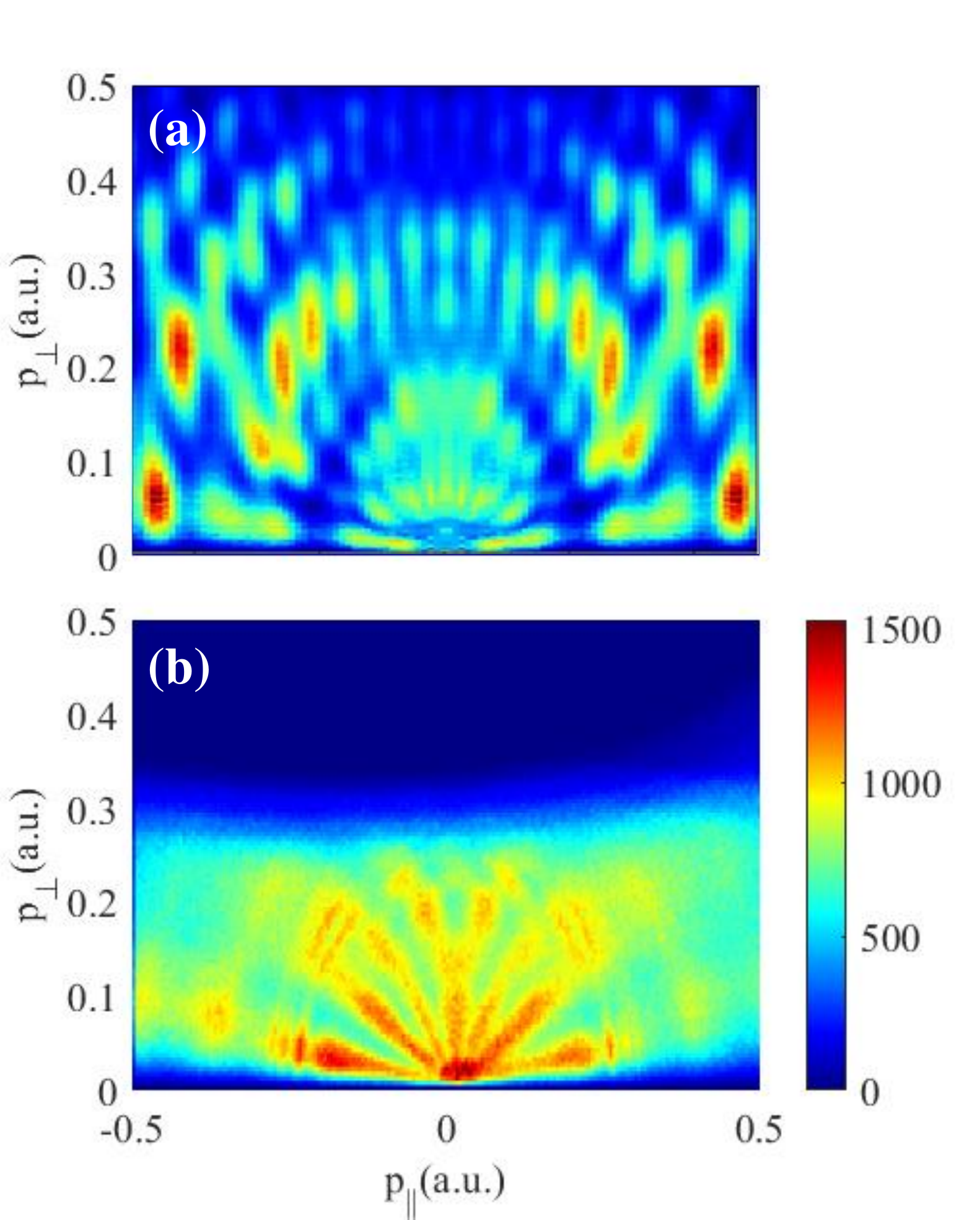}%
\caption{\label{fig:Pz_Pr}}Longitudinal (p$_{||}$=p$_{z}$) vs transverse (p$_{\perp}$=$\sqrt{p_x^2+p_y^2}$) electron momentum distribution resulting from the strong field ionisation of Ar using few cycle pulses at 0.8 PW/cm$^{2}$,  representing the intercycle and intracycle interference features. (a) theoretical simulation (b) measurement results.
\end{figure}

The photoelectron momentum distribution recorded in our experiment from the strong field ionisation of Ar at intensity 0.8 PW/cm$^{2}$ has sufficiently high resolution to resolve the low energy features. They are revealed as a superposition of multiple symmetric patterns about the polarisation axis.  In Fig. \ref{fig:Pz_Pr}, we can see that the higher energy region of the spectrum contains a series of intercycle features in the form of ATI rings, whereas intracycle fan-like stripes emerging radially outwards in the range $-0.3 \:< p_{\parallel}< \:0.3$ can be seen in the low energy regions. The low energy features do not exactly match in Fig. \ref{fig:Pz_Pr} (a) and (b),  as the intensity could be slightly different in experiment and simulation, and volume effects which are not taken into account in the simulation.

\subsection{Momentum/Kinetic energy dependence}
\begin{figure}[h!]
\includegraphics[scale=0.54]{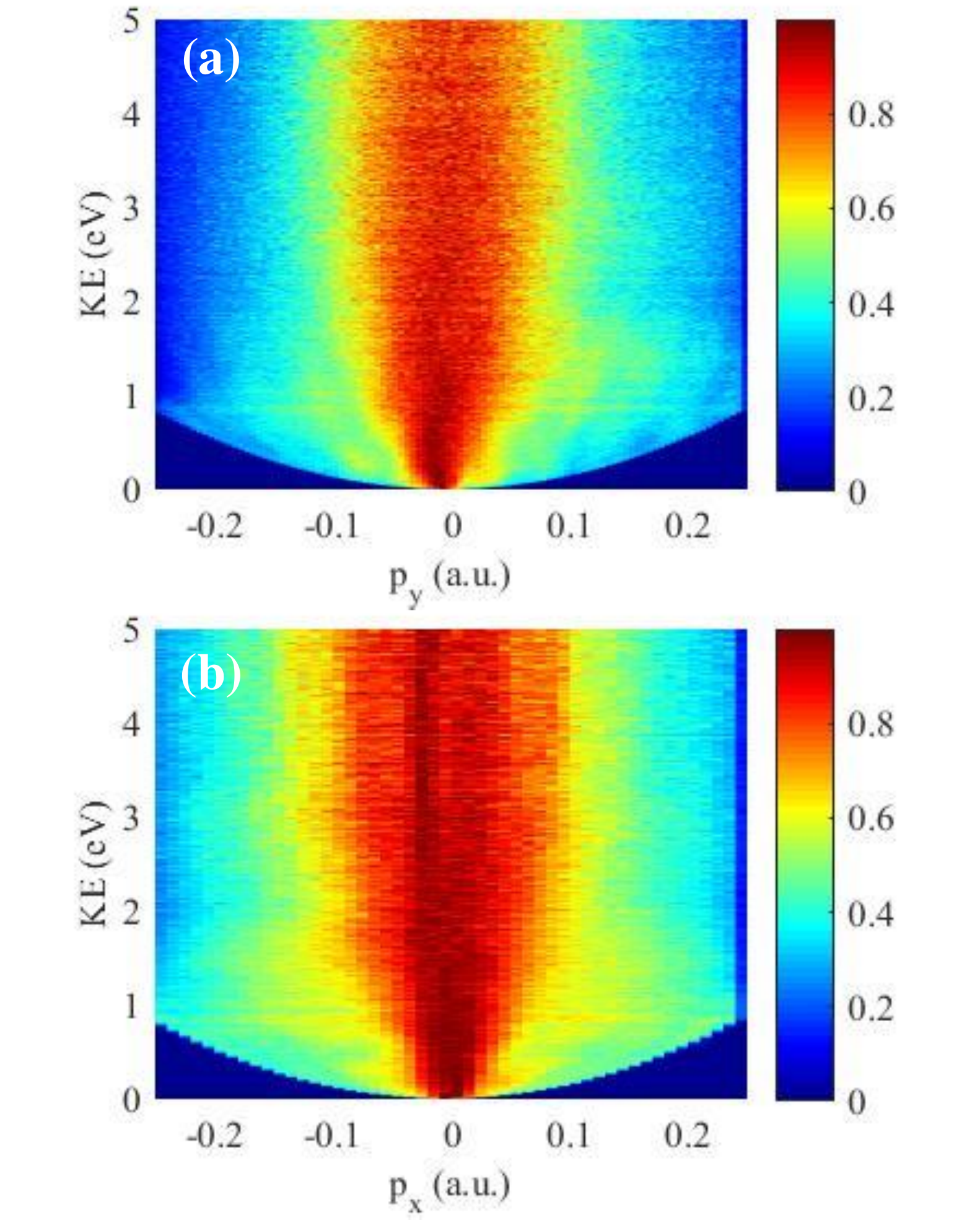}%
\caption{\label{fig:PxPyVsKE}}Energy resolved transverse electron momentum spectrum at 0.8 PW/cm$^{2}$. (a) low energy electrons for transverse momentum along the laser propagation direction experience a tilt in the negative direction and then in the positive direction with increasing kinetic energy. (b) no tilting in the low energy region for the transverse momentum perpendicular to the laser propagation direction, but a spread can be observed at higher kinetic energy.
\end{figure}

The influence of relativistic nondipole effects on the low energy photoelectrons can be seen in the plot of transverse momentum $P_y$ along the laser propagation direction vs photoelectron energy, showing the tilting for very low energy photoelectrons (0--0.5 eV). This tilt is absent for the transverse momentum $P_x$ perpendicular to the laser propagation direction (see Fig. \ref{fig:PxPyVsKE}).

The  photoelectrons experience strong Coulomb focusing effects due to the rescattering electron trajectories that correspond to very low energy. As the low energy photoelectrons typically get ionised at the peak of the laser field and return to the parent ion with close to zero momentum, therefore they stay longer around the ion and experience stronger Coulomb attraction. The interaction between the Coloumb focusing and relativistic nondipole effects leads to the spectral narrowing of the low energy TEMD with the peak shift in the negative  direction. However, the high-energy TEMD gets broadened with a forward peak shift. This energy-dependent offset can be used as a self-referencing technique for the detection and analysis of the relativistic nondipole effect, without relying on the auto-ionisation of Rydberg state molecules to calibrate the transverse momentum offset \cite{Smeenk_2011, Ludwig_2014}. 

\begin{figure}[h!]
\includegraphics[scale=0.33]{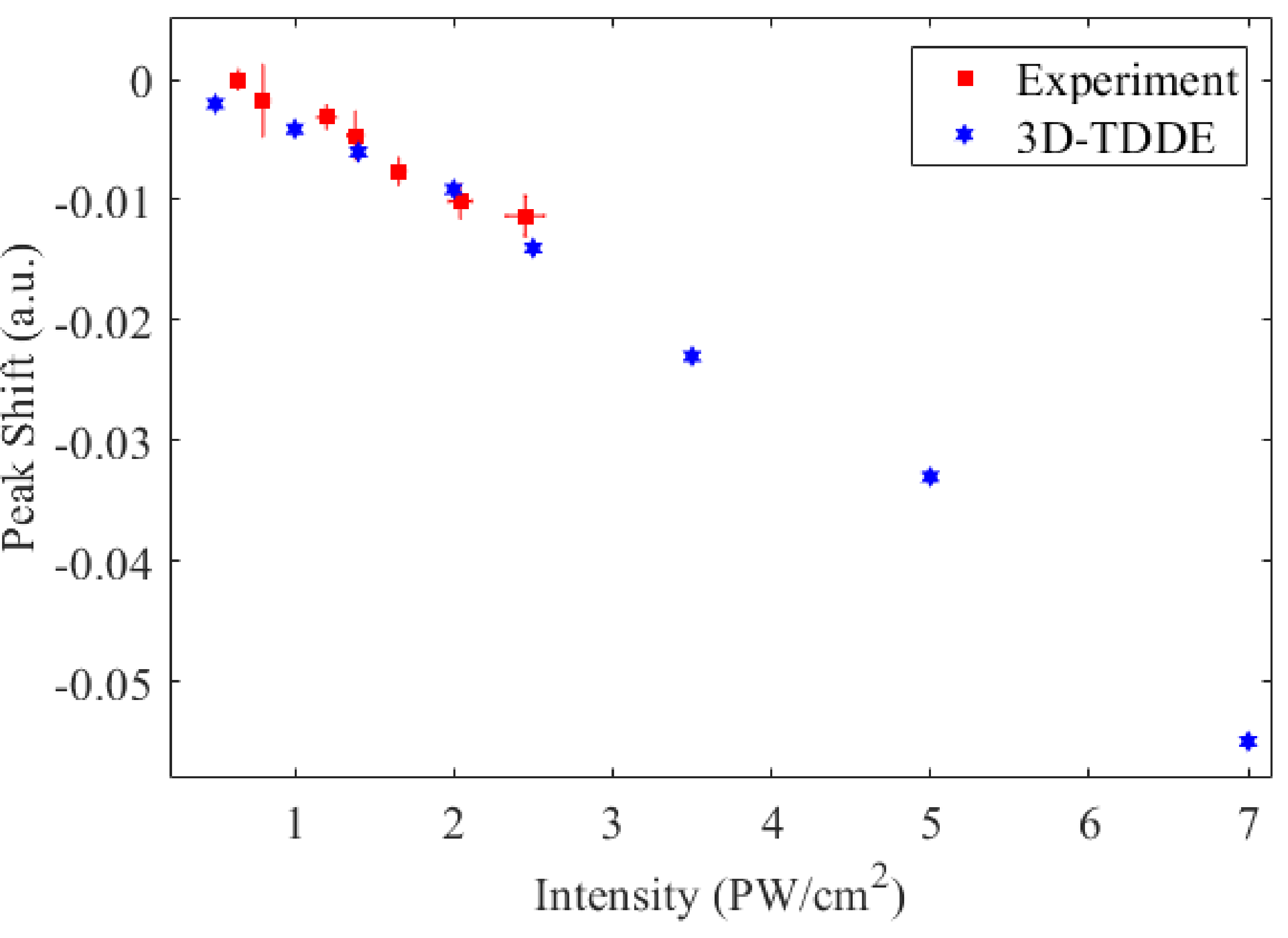}%
\caption{\label{fig:peakshiftVsintensity}} With increasing intensity, the relative peak shift of the TEMD increases in the negative direction.
\end{figure}

At the intensity 0.8 PW/cm$^{2}$, it was already expected and has been already reported that the photoelectrons show a negative peak shift. As can be seen in the results from our previous work (Fig. \ref{fig:peakshiftVsintensity}), which compares the experimental data obtained in the intensity range 0.6--3.0 PW/cm$^{2}$ with the 3D-TDDE simulation results in the intensity range 0.5--7.0 PW/cm$^{2}$, the peak corresponding to the very low energy electrons continues shifting in the negative direction with increasing intensity \cite{Haram_2019}.

\begin{figure}[h!]
\includegraphics[scale=0.38]{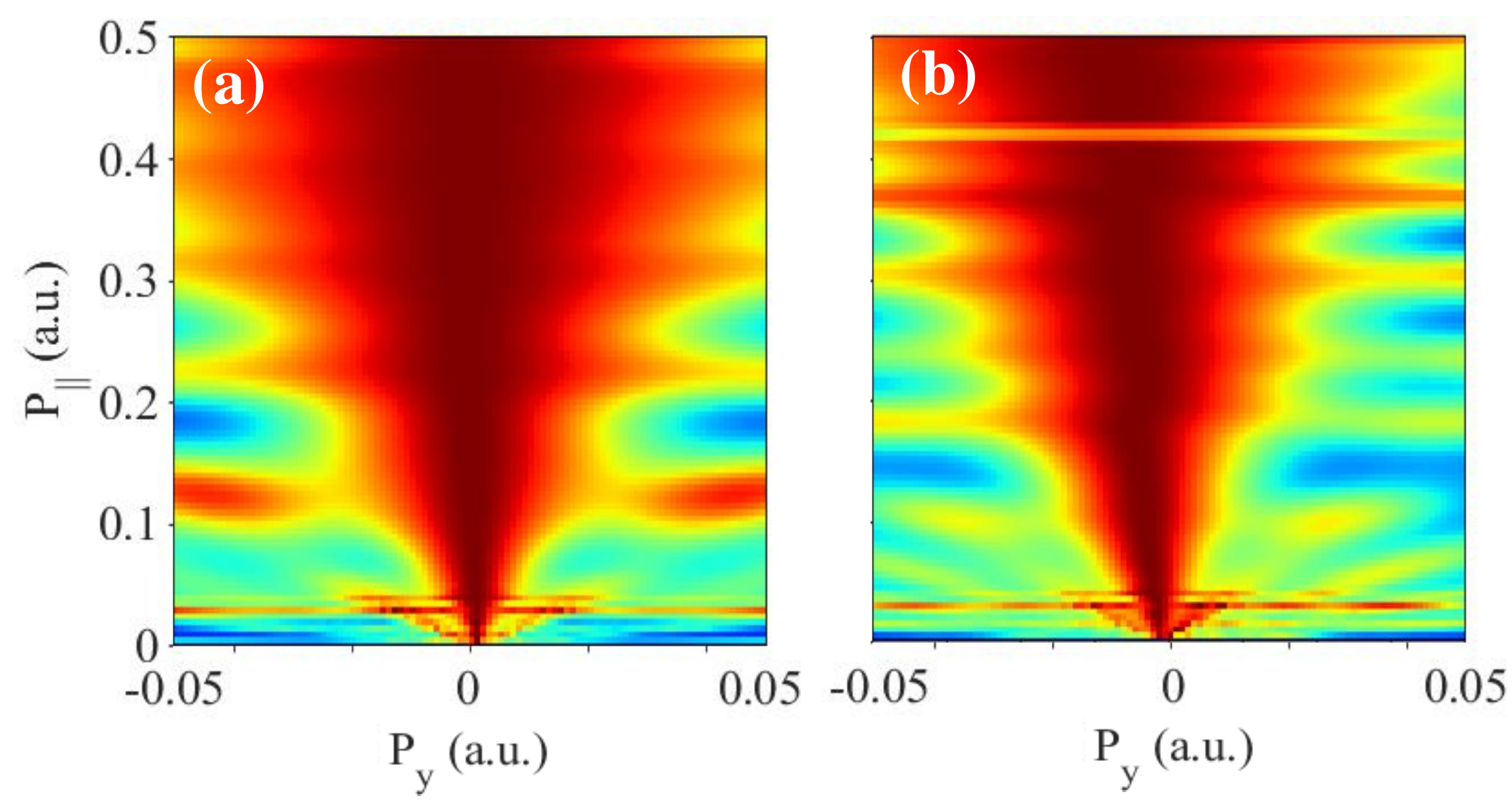}%
\caption{\label{fig:NormalisedMD_PyPz_H_L}}Simulation results: Normalised transverse vs longitudinal momentum spectra at (a) low (0.3 PW/cm$^{2}$) and (b) high (0.8 PW/cm$^{2}$) intensities.
\end{figure}

The comparison of asymmetry as observed in the simulated TEMD along the laser propagation direction at two different intensities 0.3 PW/cm$^{2}$ and 0.8 PW/cm$^{2}$ is presented in Fig. \ref{fig:NormalisedMD_PyPz_H_L}. It is evident from the figure that the momentum distribution is symmetric about zero momentum at low intensity Fig. \ref{fig:NormalisedMD_PyPz_H_L}(a). However, at a higher intensity, not only the very low energy electrons are shifted in the negative direction, but also the tilt of the photoelectron momentum distribution is quite obvious Fig. \ref{fig:NormalisedMD_PyPz_H_L}(b).

\begin{figure}[h!]
\includegraphics[scale=0.34]{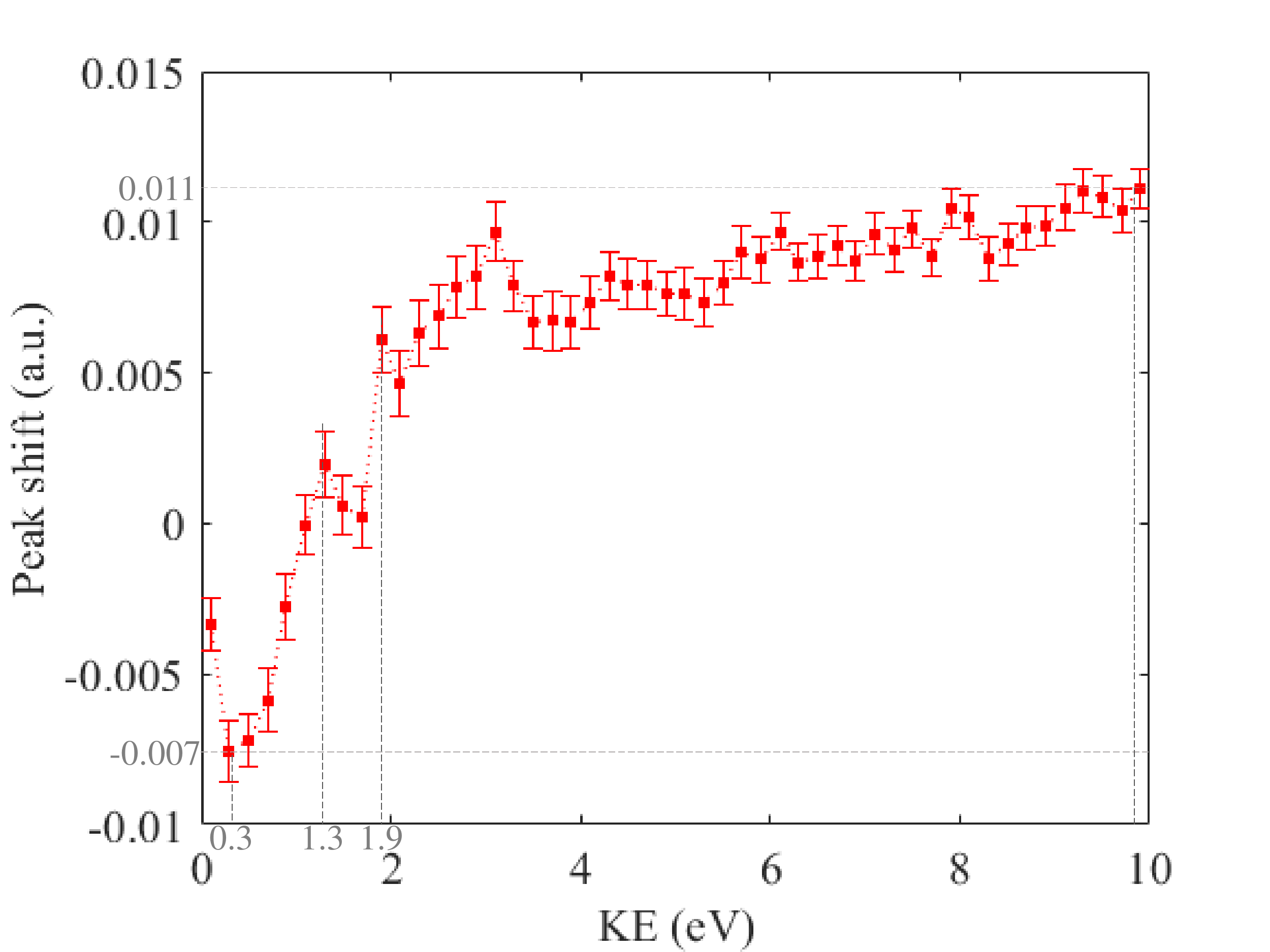}%
\caption{\label{fig:KEvsNormalisedAsymmetry}} The overall $P_{y}$ resolved peak shift is in the forward direction: 7 x 10$^{-3}$ a.u. for 0.3 eV electrons increasing to 1 x 10$^{-2}$ a.u. for 10 eV electrons. However, for very low energy electrons the peak shift is in the negative direction, which reverses its direction for the electrons greater than 0.3 eV and then follows oscillatory behaviour with increasing energy after 1.9 eV.
\end{figure}

The quantitative analysis of this tilt is performed by determining the peak position of the transverse momenta $P_y$ along laser propagation direction, integrated over the energy range 0.2 eV as a function of photoelectron energy by employing the same procedure as used in \cite{Haram_2019}. As shown in the Fig. \ref{fig:KEvsNormalisedAsymmetry}, the zero energy or close to zero energy electrons show a negative peak shift i.e. -3 x 10$^{-3}$ a.u., which keeps on increasing in the negative direction for the photoelectrons having energy less than 0.3 eV. The cause of this negative shift is the strong Coulomb focusing effect for electrons having close to zero energy. However, the very low energy photoelectrons in the range 0.3--1.3 eV experience a push in the forward direction because of the radiation pressure, causing a net peak shift of 6 x 10$^{-3}$ a.u.. With increasing energy, instead of continuing forward shift the photoelectrons experience a negative peak shift again. The shift pattern continues to oscillate from negative--positive--negative shift for the photoelectron having energy greater than 1.9 eV but in a smaller interval of approximately 1 x 10$^{-3}$ a.u.. The overall
peak shift, however, keeps on increasing from 7 x 10$^{-3}$ a.u. to 1 x 10$^{-2}$ a.u..

\begin{figure}[h!]
\includegraphics[scale=0.44]{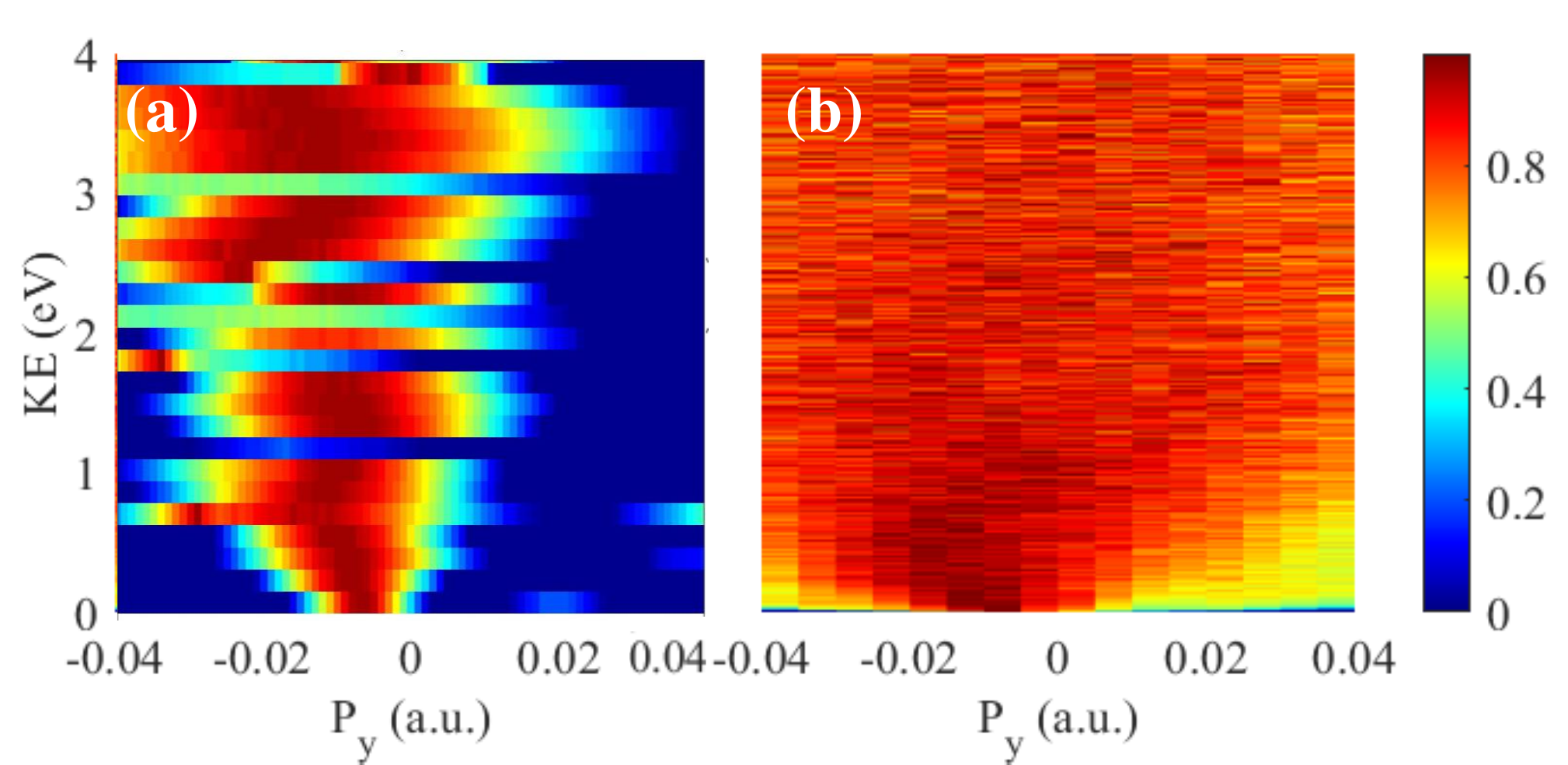}%
\caption{\label{fig:MD_PyKE} Energy resolved transverse electron momentum spectra: (a) simulation results at 1.4 PW/ cm$^{2}$, (b) experimental results at 0.8 PW/ cm$^{2}$. }
\end{figure}

The same trend of the peak shift has been observed for the simulation results obtained at even higher intensity. Fig. \ref{fig:MD_PyKE} compares the experimental data obtained at 0.8 PW/cm$^{2}$ with the simulation results obtained at 1.4 PW/cm$^{2}$, which reflects that the overall shift might be different as a function of intensity. However, close to zero energy and the very low energy electrons, the main reason of negative peak shift is due to Coulomb focusing and hence they behave the same way in this intensity range.

\begin{figure}[h!]
\includegraphics[scale=0.34]{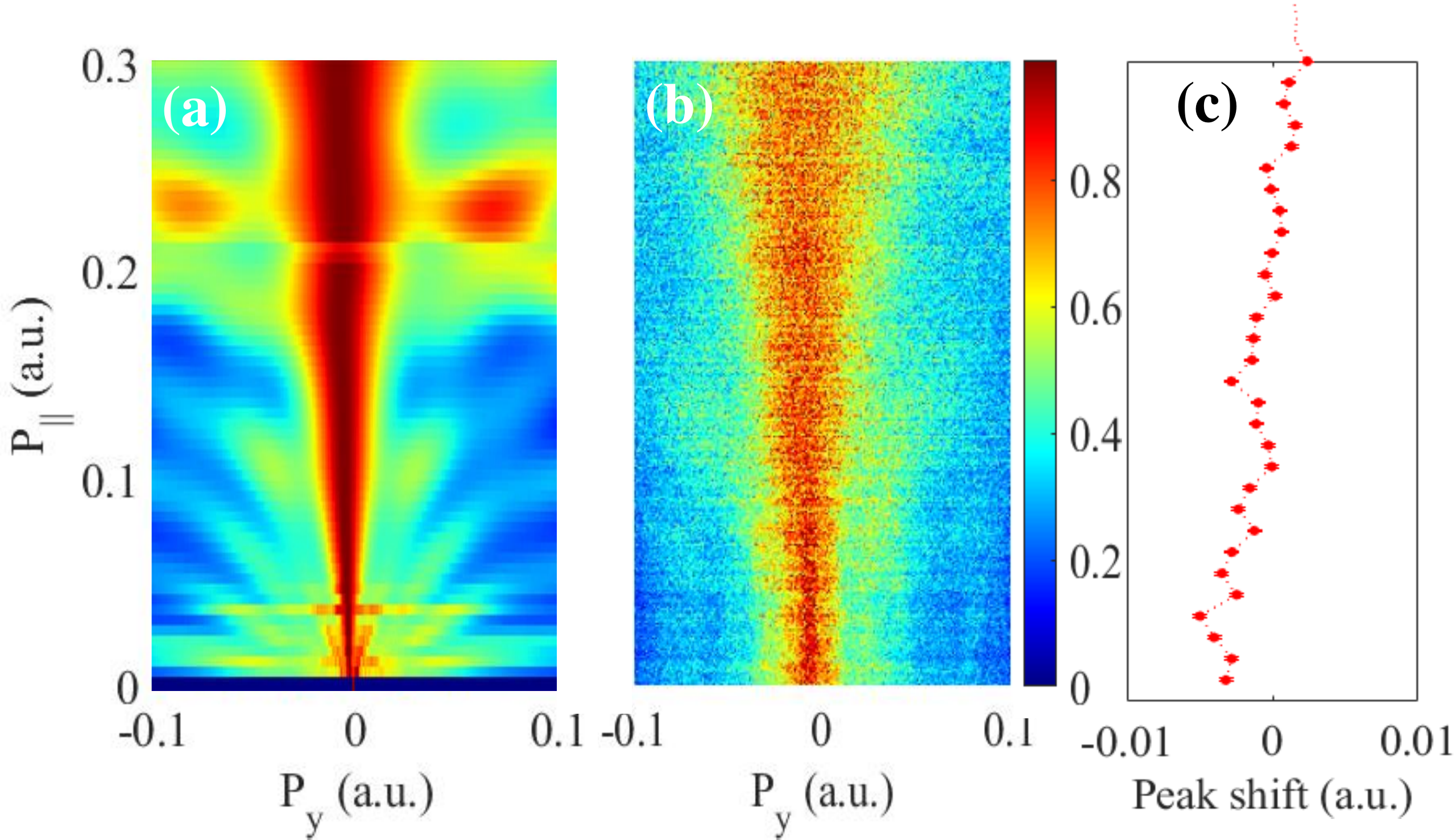}%
\caption{\label{fig:NormalisedMD_PyPz_zoom}}Normalised transverse vs longitudinal momentum spectra at high intensity 0.8 PW/ cm$^{2}$. Comparison of (a) simulation and (b) experimental results. (c) The plot of extracted peak shift from the experimental data.
\end{figure}

To further explore the tilt caused by the relativistic nondipole effects on the low energy photoelectrons in Fig. \ref{fig:PxPyVsKE}, we plot transverse momentum in the laser propagation direction vs longitudinal momentum. Fig \ref{fig:NormalisedMD_PyPz_zoom} presents the comparison between the simulated 
spectrum and  measured TEMD at 0.8 PW/cm$^{2}$. An overall tilt in P$_{y}$ in both (a) simulated spectra and (b) measured spectra is observed. To highlight the features of interest, we zoom in the transverse momentum in a narrow range [-0.05 a.u., 0.05 a.u.]. and extract the peak position of the transverse momentum P$_\perp$ integrated over the momentum range 0.02 a.u.. as shown in the (c) plot. The peak shift is extracted by the same procedure as described in \cite{Haram_2019}. The peak shift is determined by the asymmetry in the cusp of the TEMD. For this purpose, the function $V(P_{\perp}) \equiv \ln {W(P_{\perp})}$ is analysed in a narrow range of transverse momenta $|P_{\perp}|\le 0.1$ a.u., where  $W(P_{\perp})$ is the ionisation rate. The experimental data is fitted with the function $V(P_{\perp})= B + A|P_{y} - \beta|^\alpha$ by performing a series of least square fits, where $A$, $B$, $\alpha$ and $\beta$ are the fitting parameters with $\beta$ accountable for the peak shift $\left\langle P_{\perp}\right\rangle$. The same procedure has been employed to determine the $P_y$ resolved peak shift in Fig. \ref{fig:KEvsNormalisedAsymmetry}.

For zero and close-to-zero energy electrons the peak shift is negative, which keeps on oscillating between negative and positive direction with increasing longitudinal momentum but shows an overall positive shift i.e. a shift from - 3 x 10$^{-3}$ a.u. to 4 x 10$^{-3}$ a.u.. This oscillatory behavior of the peak shift can be attributed to the photoelectrons with different trajectories.
%\begin{figure}[h!]
%\includegraphics[scale=0.36]{NormalisedMD_PyPz.pdf}%
%\caption{\label{fig:NormalisedMD_PyPz}}Normalised transverse vs longitudinal momentum spectra at high intensity, comparison of (a) simulation and (b) experimental results.
%\end{figure}

\begin{figure}[h!]
\includegraphics[scale=0.48]{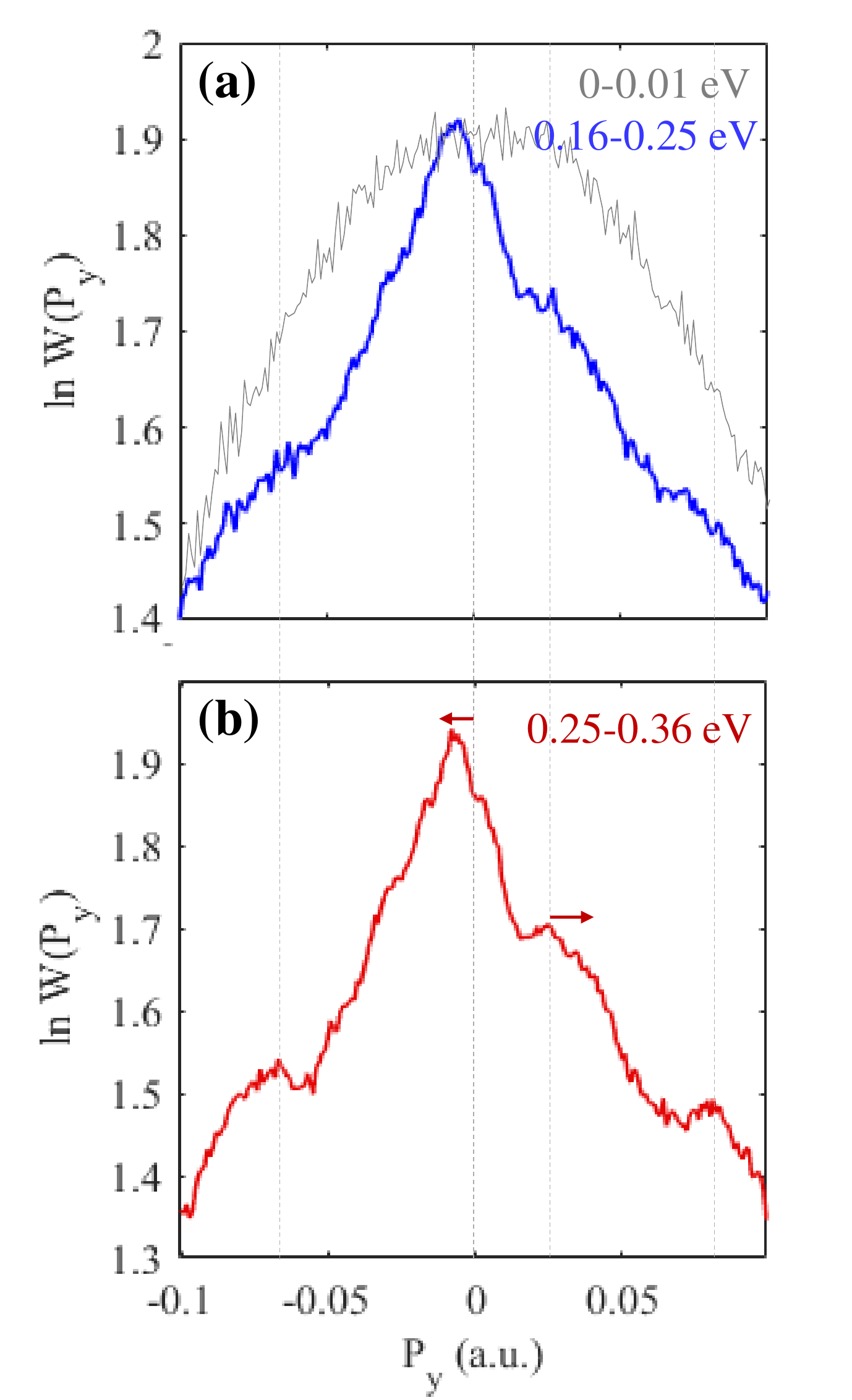}%
\caption{\label{fig:Py_yield}} Measured TEMD along the laser propagation direction obtained  0.8 PW/cm$^{2}$. (a) Wings of TEMD start to appear for the electrons in the energy range 0.16-0.25 eV with an asymmetric distribution (blue) compared to the distribution for very low energy electrons (grey). (b) More pronounced wings of the TEMD for the electrons in the energy range 0.25-0.36 eV. The central peak is shifting in the negative direction whereas wings tend to shift in the positive direction.
\end{figure}

Further investigation on the TEMD along the laser propagation direction for the low energy electrons (0-1 eV) integrated over the momentum range 0.1 a.u. obtained at 0.8 PW/ cm$^{2}$ shows quite interesting behaviour. For close to zero energy electrons (0-0.01 eV) , the  TEMD seems to be symmetric and uniform. However, the wings of the TEMD start to appear progressively with increasing energy in the range 0.15-0.4 eV, which gradually fade away leading to a broader distribution after this energy range. The central peak of the TEMD shifts in the negative direction for the electrons in the energy range 0.16-0.25 eV as shown in Fig \ref{fig:Py_yield} (a). These are the low energy electrons with long trajectories that may revisit 
the parent ion \cite{Chelkowski_2015}. This peak shift keeps on increasing in the negative direction until the electrons gain energy of $\sim$ 0.36 eV. In contrast, the wings of the electrons are shifted in the positive direction, since they are formed by the direct electrons which never get a chance revisit the parent ion (see Fig \ref{fig:Py_yield} (b)). The high energy electrons with energy greater than 0.36 eV  corresponding to short trajectories contribute to the overall positive peak shift.

\subsection{CEP dependence}

\noindent\textit{CEP averaged momentum distributions}\\

\begin{figure}[h!]
\includegraphics[scale=0.43]{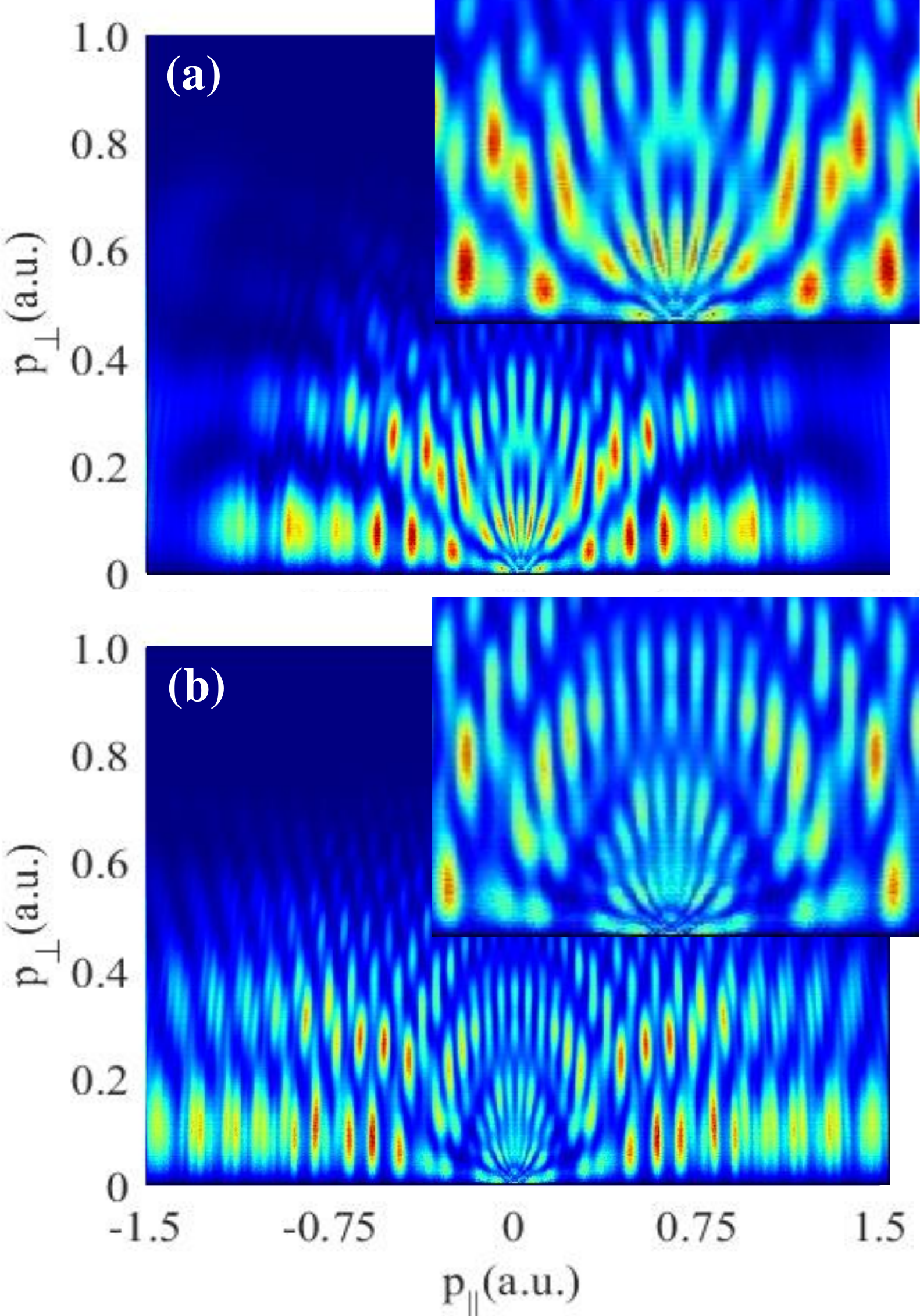}%
\caption{\label{fig:MD_PzPr_CEPaveraged}}CEP averaged photoelectron momentum distributions. (a) at low intensity 0.3 PW/cm$^{2}$, (b) at high intensity 0.8 PW/cm$^{2}$.
\end{figure}

The simulated CEP averaged photoelectron momentum distributions based on 3D-TDDE at  0.3 PW/cm$^{2}$ and 0.8 PW/cm$^{2}$ are as shown in Fig. \ref{fig:MD_PzPr_CEPaveraged} (a,b). In general, these spectra exhibit a number of features, mainly consisting of intercycle and intracycle interferences. \textit{Intercycle interferences} cause the momentum distribution to evolve into concentric ATI-rings centered at zero momentum and are formed as a result of superposition of electron trajectories corresponding to complex release time during multiple optical cycles. These rings are insensitive to the Coulomb potential and their yield depends upon the direction of photoelectron emission \cite{Arbo_2006,ARBO_2012}. Regular carpet-like features are created by the interference of ATI rings that are separated by a number of cycles \cite{Korneev_2012}. \textit{Intracycle interferences} lead to the formation of fan-like stripes caused by the interference of electron trajectories within one optical cycle. These include the direct photoelectron trajectories and the trajectories corresponding to those electrons which are deflected by the Coulomb potential without experiencing any hard recollisions \cite{Lai_2015}. The intracycle interferences are also responsible for the coherent superposition of distinct states known as Freeman resonances. These resonant features are formed as a result of a range of intensities attributed to the focal volume below the peak intensity \cite{Freeman_1987}. The pattern of these fan-like features exhibit strong dependence on the wavelength since it influences the states to be populated with a particular angular quantum number ($l$) \cite{Marchenko_2010}. The fine holographic features beside the main features emerge from the intracycle interferences between the direct photoelectron trajectories and the trajectories that correspond to hard recollisions \cite{Maxwell_2017}.

It has been established that the features in the photoelectron momentum distributions are sensitive to the laser wavelength and intensity \cite{Marchenko_2010,Marchenko_2011}. Since, we are comparing the momentum distributions at two quite different intensities, the number of ATI-rings and fan-like stripes as well as their patterns are found to be significantly different (see Fig. \ref{fig:MD_PzPr_CEPaveraged}). Considering these momentum distributions are CEP averaged, the features are found to be symmetric about zero momentum.\\

\noindent\textit{CEP dependent momentum distributions}\\

\begin{figure}[h!]
\includegraphics[scale=0.32]{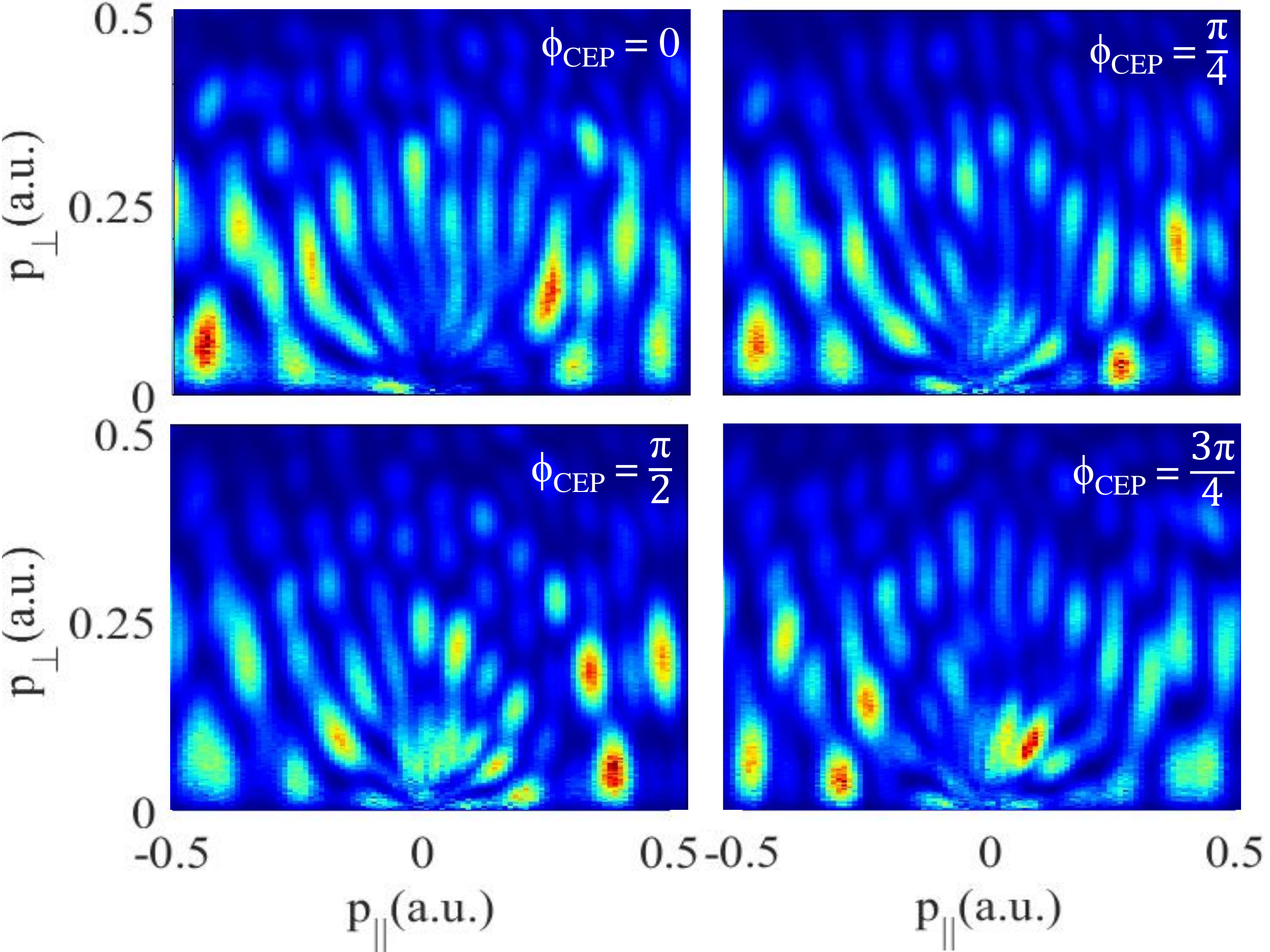}%
\caption{\label{fig:MD_PzPr_CEPdependent_low}}CEP dependent photoelectron momentum distributions at low intensity 0.3 PW/cm$^{2}$.
\end{figure}

\begin{figure}[h!]
\includegraphics[scale=0.32]{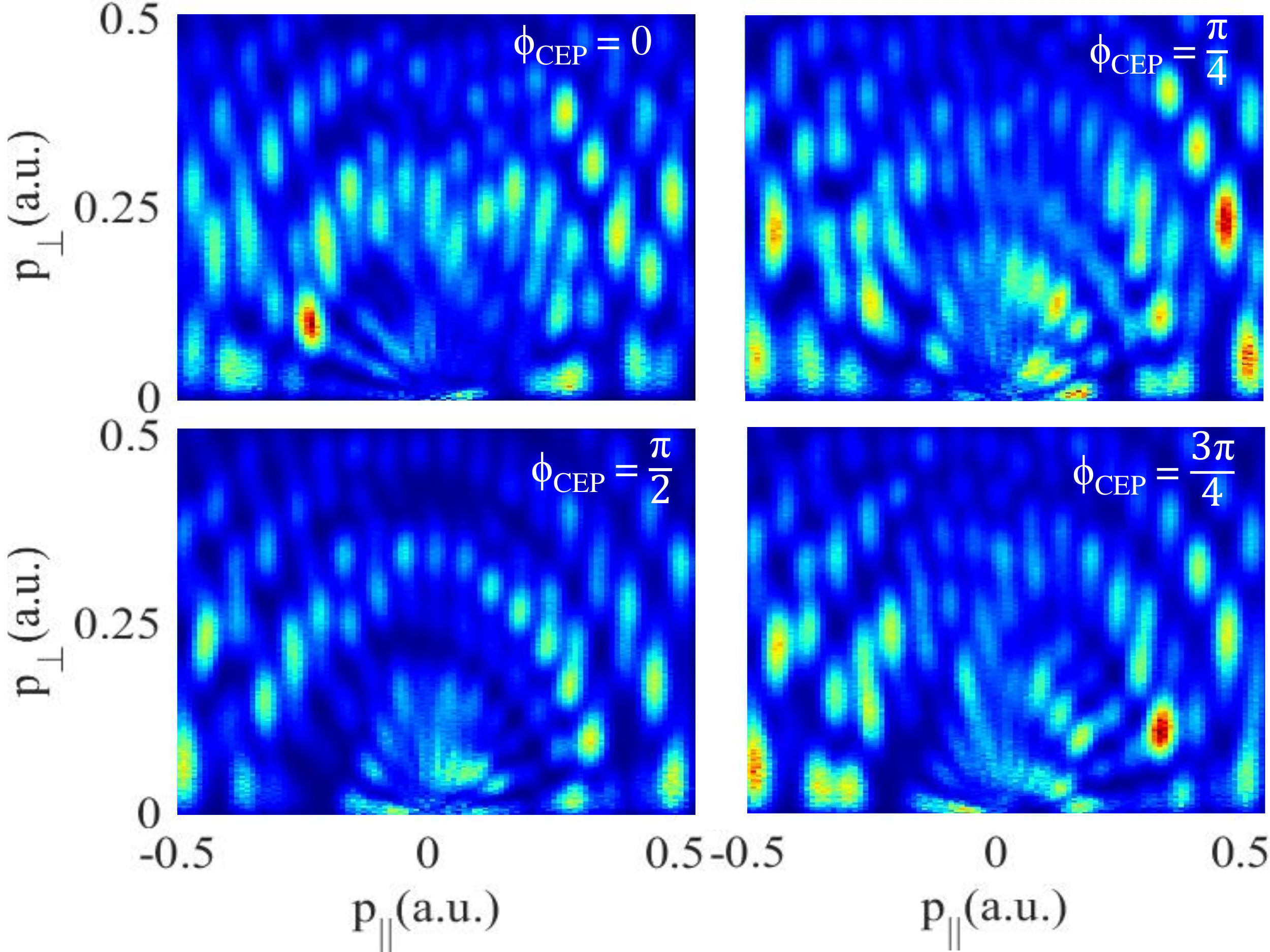}%
\caption{\label{fig:MD_PzPr_CEPdependent_high}}CEP dependent photoelectron momentum distributions at high intensity 0.8 PW/cm$^{2}$.
\end{figure}

The CEP has profound effect on the photoelectron momentum distributions. The spectra obtained at two different intensities show quite unique features for each CEP as shown in Fig. \ref{fig:MD_PzPr_CEPdependent_low} and Fig. \ref{fig:MD_PzPr_CEPdependent_high}. 

\begin{figure}[h!]
\includegraphics[scale=0.34]{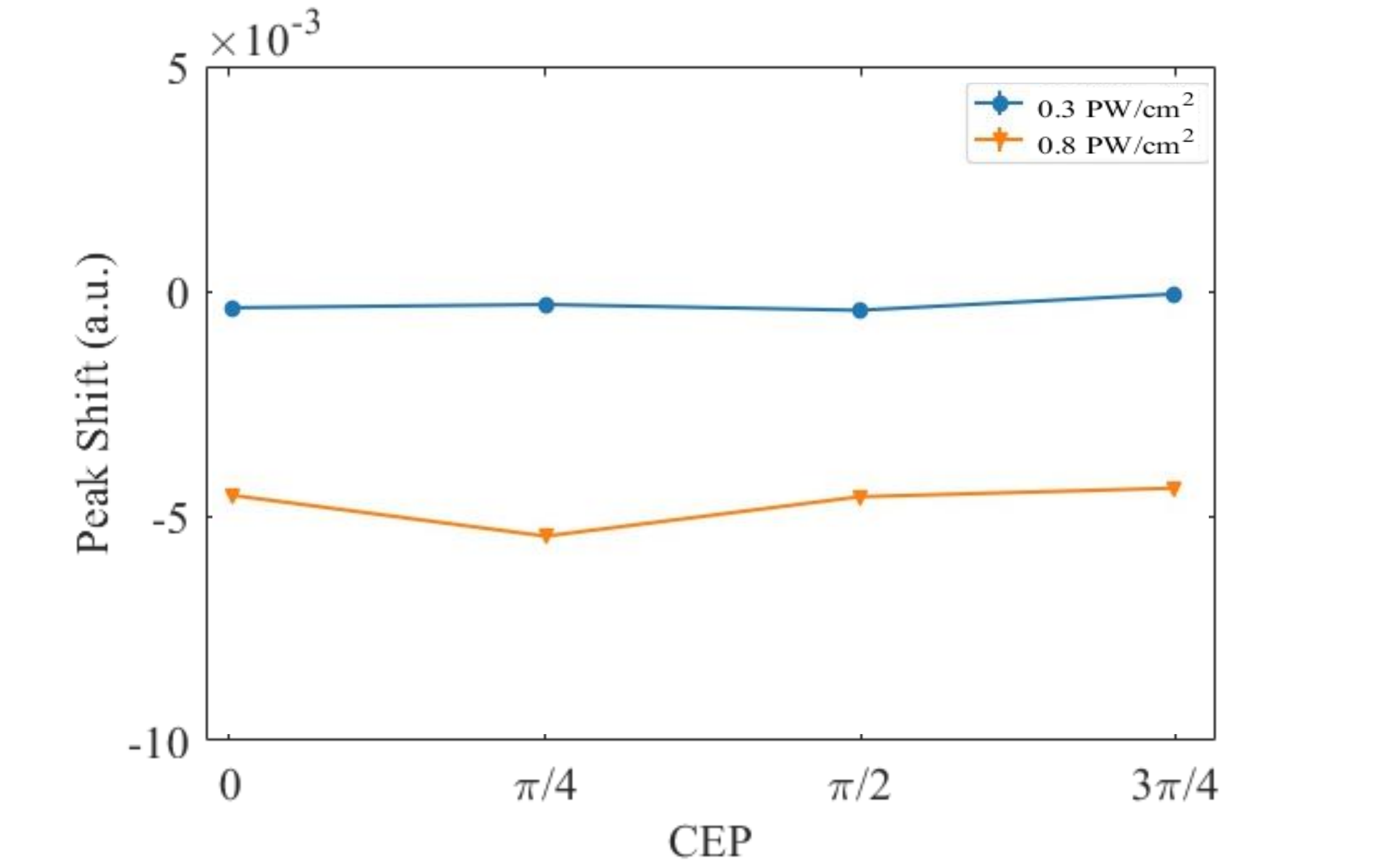}%
\caption{\label{fig:CEP_peakshift}}CEP dependent peak shift of the TEMD along laser propagation direction at 0.3 PW/cm$^{2}$ and 0.8 PW/cm$^{2}$.
\end{figure}

The features are significantly different for different CEP values i.e. not alike nor symmetric, since a specific CEP value may enhance the population of only a particular resonant feature. However, as reported earlier in our previous paper, the effect of CEP on the nondipole effects is not resolvable within experimental uncertainty. In order to explore these effects, the peak shift of the TEMD along the laser propagation direction was extracted for the sine and cosine pulses, which turned out to be the much smaller than the experimental uncertainty \cite{Haram_2019}. Further investigation based on the simulation results at 0.3 PW/cm$^{2}$ and 0.8 PW/cm$^{2}$ supports the previous results showing that for different CEP values the peak shift is extremely small especially at lower intensity (see Fig. \ref{fig:CEP_peakshift}).

\section{Classical Analysis}

To investigate the role of photoelectron trajectories on the relativistic nondipole effects, a simple physical picture based on a modified three step model is presented. In the first step, the electron is ionised close to the peak of laser electric field. In the next step, the photoelectron is accelerated by the driving laser field along its polarisation axis and follows its oscillatory motion with a constant drift motion given by the laser field at the tunnel exit. Due to this drift momentum, which the photoelectron possesses when the laser pulse is over, the photoelectron gets detected at the detector. However, perpendicular to the polarisation axis, the photoelectron is pushed along its propagation axis with a constant velocity due to the Lorentz force. In the third step, the photoelectron obtains a negative transverse momentum when it revisits and gets elastically scattered by the Coulomb potential of parent ion,  which is similar to the Rutherford scattering process. The drift momentum gained in the second step is responsible for the particular trajectories that allows the low energy rescattering and cause the photoelectron to be strongly focused. Thus, the negative photoelectron momentum shift is sensitive to the photoelectron trajectory, and it happens within one optical cycle. \\

\noindent\textit{Classical electron trajectory model:}\\

The validity of this modified three step model can be confirmed by performing a pure classical electron trajectory simulation for different ionisation times. Experimentally, it can be confirmed by using CEP locked few-cycle pulse or two-color laser field to break the top/bottom symmetry, which is the subject of our next work. In this way, the photoelectron trajectory and its momentum shift along laser propagation direction are expected to be different for P$_{||}>0$ and P$_{||}<0$. 

Here, we have performed a full classical electron trajectory simulation. The pump laser is propagating along y-axis and its electric field component is linearly polarised along z-axis:
\begin{equation}
    \begin{pmatrix}
    E_x \\ 
    E_y \\ 
    E_z
    \end{pmatrix}
    =
    \begin{pmatrix}
    0\\
    0\\
    E_0\;sin^2(\frac{\pi (t-\frac{y}{c})} {\tau_0}) \; cos[\omega(t-\frac{y}{c})+ \phi_0]
    \end{pmatrix}
    \label{eq:1}
\end{equation}

\noindent where $E_0$ is the peak electric field strength (0.06 a.u.), $\tau_0$ is the pulse duration (5 optical cycle), $\omega$ is the pump laser carrier frequency, $\phi_0$ is the carrier-to-envelope phase (CEP), and c is the speed of light in vacuum. The magnetic field component of the pump laser is linearly polarised along x-axis:
\begin{equation}
    \begin{pmatrix}
    B_x \\ 
    B_y \\ 
    B_z
    \end{pmatrix}
    =
    \begin{pmatrix}
    -E_z\\
    0\\
    0
    \end{pmatrix}
    \label{eq:2}
\end{equation}

The dynamics of electron in the presence of pump electric and magnetic field, as well as the Coulomb potential force from the ion is given by:
\begin{equation}
    \begin{pmatrix}
    F_x\\
    F_y\\
    F_z
    \end{pmatrix}
    =
    \begin{pmatrix}
    - \frac{dU(r)}{dr} \frac{x}{r}\\
    \frac{B_x}{c} v_z - \frac{dU(r)}{dr} \frac{y}{r}\\
    -e E_z - \frac{B_x}{c} v_y- \frac{dU(r)}{dr} \frac{z}{r}
    \end{pmatrix}
    \label{eq:3}
\end{equation}
\noindent where $U(r) = -1/\sqrt{r^2 + s}$ is a softcore Coulomb potential with $s = 0.1 \; a.u.$ , $r = \sqrt{x^2+y^2+z^2}$ is the distance between electron and ion.
\begin{figure}[h!]
\includegraphics[scale=0.42]{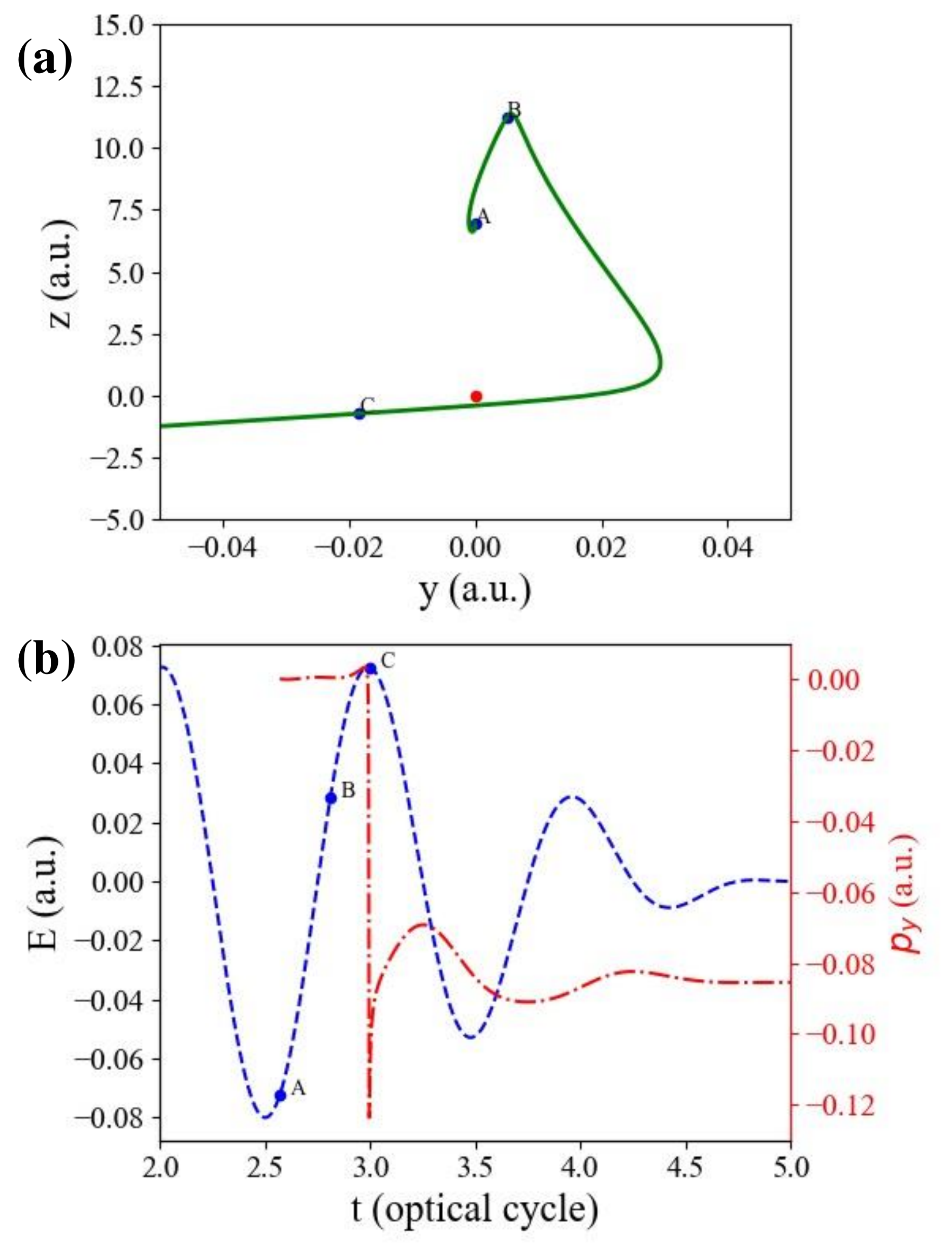}%
\caption{\label{fig:classicalTrajectory_negative}(a) The calculated typical electron trajectory with negative momentum along laser propagation direction, where the initial exit position $z_0 = 6.97 \; a.u.$, initial velocity $vz_{0} = -0.184\; a.u.$ and $t_0 = 2.57\; o.p.$ is used in the classical simulation. (b) The time evolution of the driving electric field (blue dashed curve) and time dependent electron momentum $P_y$. The point A, B, C in (a) indicates the electron position when $t = 2.57\; o.p.$ (point A in (b)), $t = 2.81 \;o.p.$ (point B in (b)),  and $t = 2.996 \;o.p.$ (point C in (b)) respectively.}
\end{figure}
$v_x,v_y,v_z$ are the three electron velocity components. Given initial conditions, including electron tunneling time $t_0$, exit site $(0,0,z_0)$ and assuming zero initial velocity after tunneling ionisation, the final 3D momentum and position of the free electron can be calculated by solving the Newton function (\ref{eq:3}), where an integration time of 7 optical periods (o.p.) is used. By scanning the tunneling time $t_0$ and exit site position $z_0$, we found the negative peak shift of the TEMD that corresponds to the electrons having tunneling time at around the local peaks of the driving electric field and the exit site around $z_0 = 6.97 \; a.u.$.

\begin{figure}[h!]
\includegraphics[scale=0.42]{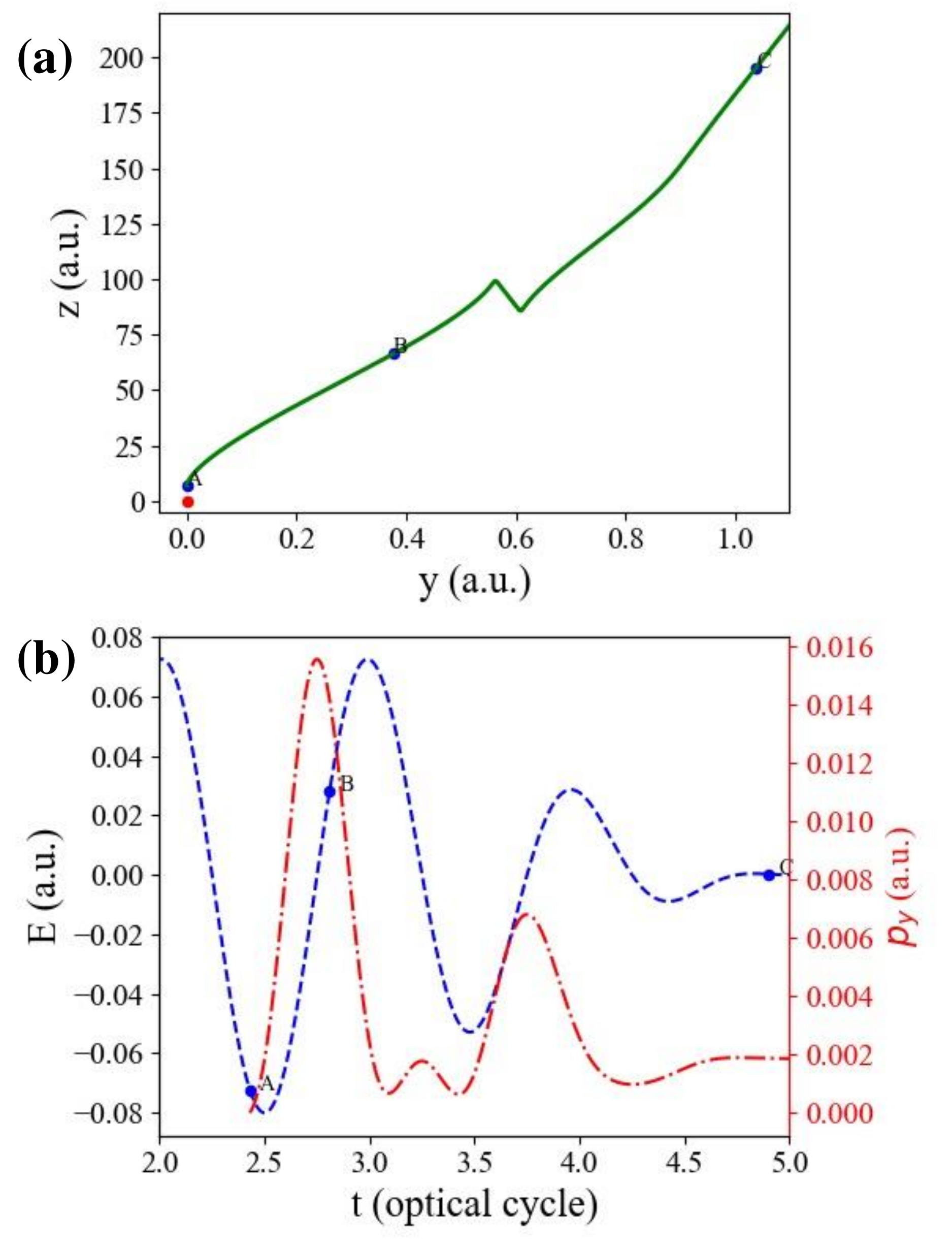}%
\caption{\label{fig:classicalTrajectory_positive}}Same as Fig. 2, but with ionisation time set as $t_{0} = 2.43\;o.p.$, the exit site as $(x_{0} = 0,y_{0} = 0, z_{0} = 6.97\; a.u.)$, and exit velocity as $(vx_{0} = 0,vy_{0} = 0, vz_{0} = 0.184\; a.u.)$. The point A, B, C in (a) indicates the electron position when $t = 2.43\; o.p.$ (point A in (b)), $t = 2.81\; o.p.$ (point B in (b)),  and $t = 4.9\; o.p.$ (point C in (b)) respectively.
\end{figure}

Fig. \ref{fig:classicalTrajectory_negative} shows a typical electron trajectory with negative momentum shift along y-axis, where the electron is ionized at $t_{0} = 2.57 \; o.p.$ (electric field is peaked at 2.5 o.p.), the exit site is $(x_{0} = 0,y_{0} = 0, z_{0} = 6.97\; a.u.)$, and exit velocity is $(vx_{0} = 0,vy_{0} = 0, vz_{0} = -0.184 \; a.u.)$ Within half optical cycle, the electron is driven away from the parent ion and reaches largest ion/electron distance at $t = 2.81\; o.p.$. After that, the electron is driven back towards the ion, and gains a positive spatial displacement along laser propagation direction (y-axis) due to the radiation pressure produced by the laser field. At t = 3.0, the electron is re-scattered by the parent ion and gains a negative momentum which is opposite to the laser propagation direction. An electron trajectory with positive momentum along y-axis is possible for electrons which have not experienced re-scattering. Figure \ref{fig:classicalTrajectory_positive} shows a typical electron trajectory with positive momentum shift the electron is no longer driven back to its parent ion, the electron can gain positive momentum along y-axis.

\section{Conclusion and Outlook}
We have investigated how the low-energy features of the photoelectron momentum and energy spectra are impacted by the relativistic nondipole effects for ultrashort linearly polarised near-infrared laser field. The measured and simulated photoelectron momentum spectra for Ar show very interesting sharp features in the form of ATI-rings and Freeman resonances due to the inter- and intra-cycle resonances. Further investigations
on the energy-resolved relativistic nondipole effects reveal an asymmetry or tilt in both momentum and energy spectra. It shows that the central peak shift for the low-energy electrons is always in the negative direction with reference to the laser propagation direction. However, the high energy electrons that also form the wings of the TEMD contribute towards the positive shift. Moreover, this peak shift is found to be oscillating between negative and positive direction, which can be ascribed to specific electron trajectories as explored by classical analysis based on modified three step model. The influence of CEP on the relativistic nondipole effects has also been investigated, showing that it has a very small effect on the peak shift of the TEMD such that it is difficult to resolve it within experimental uncertainty.

\section*{Acknowledgement}
This research  was supported by the Australian Research Council
Discovery Project DP110101894. We also acknowledge support from the Institute for Basic Science, Gwangju, Republic of Korea, under IBS-R012-D1. N.H., A.A., U.S.S. and D.C. are supported by Griffith University International Postgraduate Research Scholarship (GUIPRS). H.X. is supported by an ARC Discovery Early Career Researcher Award DE130101628.

N.H. and H.X. contributed equally to this work.
\bibliography{PRA}

\end{document}